\documentclass{vgtc}                          % final (conference style)
\ifpdf%                                % if we use pdflatex
  \pdfoutput=1\relax                   % create PDFs from pdfLaTeX
  \usepackage{graphicx}                % allow us to embed graphics files
  \DeclareGraphicsExtensions{.pdf,.png,.jpg,.jpeg} % for pdflatex we expect .pdf, .png, or .jpg files
\else%                                 % else we use pure latex
  \usepackage{graphicx}                % allow us to embed graphics files
  \DeclareGraphicsExtensions{.eps}     % for pure latex we expect eps files
\fi%

\graphicspath{{figures/}{pictures/}{images/}{./}} % where to search for the images

\usepackage{microtype}                 % use micro-typography (slightly more compact, better to read)
\PassOptionsToPackage{warn}{textcomp}  % to address font issues with \textrightarrow
\usepackage{textcomp}                  % use better special symbols
\usepackage{mathptmx}                  % use matching math font
\usepackage{times}                     % we use Times as the main font
\usepackage{cite}                      % needed to automatically sort the references
\usepackage{tabu}                      % only used for the table example
\usepackage{booktabs}                  % only used for the table example
\usepackage[table]{xcolor}
\usepackage{colortbl}
\usepackage{amsmath}
\usepackage{amsfonts}
\usepackage{pgfplots}

\usepackage{hyperref}

\onlineid{0}

\vgtccategory{Research}

\vgtcinsertpkg

\title{PassViz: A Visualisation System for Analysing Leaked Passwords%
\thanks{This is the authors' version of the accepted paper. Please cite this paper as follows: Sam Parker, Haiyue Yuan and Shujun Li (2023) PassViz: An Interactive Visualisation System for Analysing Leaked Passwords. \emph{Proceedings of the 2023 20th IEEE Symposium on Visualization for Cyber Security} (VizSec 2023), pp.~33-42, IEEE, doi: \href{https://doi.org/10.1109/VizSec60606.2023.00011}{10.1109/VizSec60606.2023.00011}. For the published version, please visit the publisher's website via the DOI link.}}

\author{Sam Parker\thanks{e-mail: samcparker1999@gmail.com}, Haiyue Yuan\thanks{e-mail: h.yuan-221@kent.ac.uk}, Shujun Li\thanks{e-mail: s.j.li@kent.ac.uk}\\%
\scriptsize Institute of Cyber Security for Society (iCSS) \& School of Computing, University of Kent, UK}

\teaser{
  \centering
  \includegraphics[width=\linewidth]{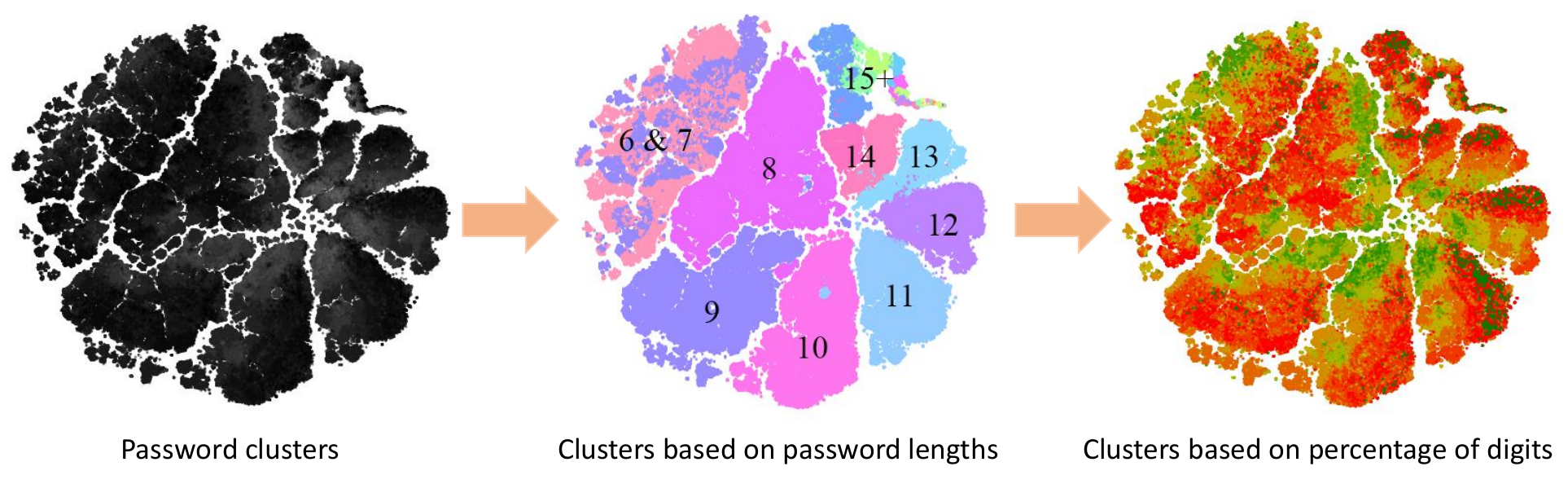}
  \caption{Examples of visualisation of different clusters for 000webhost leaked password database}
  \label{fig:teaser}
}

\abstract{
Passwords remain the most widely used form of user authentication, despite advancements in other methods. However, their limitations, such as susceptibility to attacks, especially weak passwords defined by human users, are well-documented. The existence of weak human-defined passwords has led to repeated password leaks from websites, many of which are of large scale. While such password leaks are unfortunate security incidents, they provide security researchers and practitioners with good opportunities to learn valuable insights from such leaked passwords, in order to identify ways to improve password policies and other security controls on passwords. Researchers have proposed different data visualisation techniques to help analyse leaked passwords. However, many approaches rely solely on frequency analysis, with limited exploration of distance-based graphs. This paper reports PassViz, a novel method that combines the edit distance with the t-SNE (t-distributed stochastic neighbour embedding) dimensionality reduction algorithm for visualising and analysing leaked passwords in a 2-D space. We implemented PassViz as an easy-to-use command-line tool for visualising large-scale password databases, and also as a graphical user interface (GUI) to support interactive visual analytics of small password databases. Using the ``000webhost" leaked database as an example, we show how PassViz can be used to visually analyse different aspects of leaked passwords and to facilitate the discovery of previously unknown password patterns. Overall, our approach empowers researchers and practitioners to gain valuable insights and improve password security through effective data visualisation and analysis.
} % end of abstract

\CCScatlist{
  \CCScatTwelve{Human-centered computing}{Visu\-al\-iza\-tion}{Visu\-al\-iza\-tion techniques}{Treemaps};
  \CCScatTwelve{Human-centered computing}{Visu\-al\-iza\-tion}{Visualization design and evaluation methods}{}
}

\begin{document}


% \firstsection{Introduction}

\maketitle

\section{Introduction} %for journal use above \firstsection{..} instead

Passwords are still the mostly used form of user authentication, especially for websites. Despite ongoing advancements in other forms of user authentication mechanisms, many researchers suggested that the use of passwords would continue to prevail in the foreseeable future~\cite{Herley-C2012, Bovnjak-L2019}. More recently, passwords are often used as part of a multi-factor authentication (MFA) system, where one or more other factors such as ``what you have'' (token-based) and ``who you are'' (biometric-based) authentication methods are used to provide enhanced overall security. Despite its wide use, the shortcomings of passwords such as weak passwords defined by human users are well-studied in the research literature~\cite{Herley-C2012}. One source of the weak password problem is the conflict of security and usability of passwords: stronger passwords tend to be harder to remember, and easier-to-remember passwords tend to be easier to crack~\cite{Elizabeth-S2014, Ur-B2016}. Human users tend to have different insecure behaviours around password creation, e.g., the mismatch between human users' mis-perception of a password's strength and its actual strength can lead to creation of weak passwords~\cite{Ur-B2016, Saeed-A2019}, and many users choose to reuse the same password across multiple accounts~\cite{Pearman-S2017}. Such weak passwords have led to repeated leakage of passwords from many websites, including some very large-scale incidents. The unfortunate large-scale password leaks give researchers and practitioners opportunities to study such leaked passwords to gain more knowledge and insights about how human users create passwords, in order to find better ways to refine password security controls, e.g., better password policies, password checkers and password management tools.

Most earlier password analysis work was based on simple statistics~\cite{morris1979, RIDDLE1989569}, but data visualisation has been proposed by some researchers to analyse leaked passwords~\cite{Veras-R2021, yu2016user, michael11}, utilising methods such as heat-maps, bar charts, and word clouds. To the best of our knowledge, most past studies on password visualisation are based on frequencies of passwords or segments of passwords, and only a limited number of studies~\cite{GUO-X2015, Zheng2018AnAM} investigated graph-based methods to explore structural relationships between different passwords. Different from existing solutions, this paper presents PassViz, a new graph-based data visualisation method that leverages edit distances (more precisely Levenshtein distances) and  the t-distributed stochastic neighbour embedding (t-SNE) dimensionality reduction algorithm for visualising and analysing leaked passwords in a 2-D space. We implemented PassViz as an easy-to-use command-line tool for visualising large-scale password databases, and also an interactive graphical user interface (GUI) to support interactive visual analytics of small password databases. Using the ``000webhost" leaked database as an example, we show how PassViz can be used to analyse different aspects of leaked passwords in a visually meaningful manner and also facilitate the discovery of previously unknown password patterns.

The rest of the paper is organised as follows. Section~\ref{sec:related_work} overviews some related work, followed by a detailed description of the proposed methodology given in Section~\ref{sec:methodology}. Section~\ref{sec:results} demonstrates different ways of using PassViz to conduct a visual analysis of leaked passwords in the database ``000webhost", with a discussion on the limitations of PassViz. The last section concludes this paper with future research directions.

\section{Related Work}
\label{sec:related_work}

%Despite many identified security and usability issues, passwords remain the predominant way for user authentication, and many researchers believe that it will continue to be an integral part to ensure the security of information and systems in the foreseeable future.
The understanding of password structures and patterns can provide useful insights into the password creation processes and help develop better password tools such as password strength meters~\cite{Jakobsson12}. An early attempt by Morris and Thompson~\cite{morris1979} back in the 1970s analysed 3,289 passwords and revealed some basic statistics about passwords structures, where 492 passwords can be identified in open access information sources such as dictionaries and name lists, 86\% of passwords can be categorised as one of the 6 classes (e.g., single ASCII character, four alphanumerics, and all lower cases). Similarly, an early work conducted by Riddle et al.~\cite{RIDDLE1989569} investigated 6,226 passwords for a university time-sharing system, and they discovered that user-chosen passwords are commonly based on personal information such as birthday, names or job/project related. 

Jakobsson and Dhiman~\cite{Jakobsson12} studied the relationship between the percentage of passwords' components such as words, numbers, and other special characters to establish the differences between strong and weak passwords. Differently, Taiabul Haque et al.~\cite{Taiabul-HSM2014} proposed a hierarchy of password importance that assume that users would mentally classify passwords into different levels based on the perceived importance of different sites (i.e., news portals and banking websites). By observing how users construct passwords following such a hierarchy, they uncovered that unsafe lower-level passwords can be used to crack higher-level passwords due to the behaviour of password reuse with/without modifications. In a study of empirical analysis of large-scale Chinese web passwords, Wang et al.~\cite{Wang-D2019} discovered a number of interesting password structures and semantic patterns, which are somewhat different from findings observed in English passwords. They explored 22 types of semantic information such as English names, Pinyin names, date in the format of YYYY, and date in the format of YYMMDD, which contribute to password-cracking strategies.

Leaks of real-world passwords from many websites (e.g., Yahoo, RockYou, and 12306) have become a common phenomenon these days, and they have attracted many researchers' attention to study such leaked passwords in order to gain useful insights about how human users create passwords. One group of methods for facilitating such analyses of leaked passwords is to utilise data visualisation. For instance, Bonneau et al.~\cite{Bonneau2012} collected a subset of leaked passwords from RockYou, which contain only 4-digit sequences, and another password database containing only 4-digit PINs to unlock iPhones to study the composition of 4-digit PINs. By visualising the distribution of such PINs using a heat map, they revealed that it is very likely human users choose 4-digit passwords in a format of MMDD (i.e., month-day). They concluded that birthdays have been heavily used as 4-digit passwords.  

In another work, Wang et al.~\cite{Wang2017AsiaCCS} conducted a study to compare 4- and 6-digit PINs for English and Chinese users, where heat maps were adopted to visualise date-related features in such PINs. To further explore how dates are used in the password creation process, Veras et al.~\cite{Veras2012VizSec} developed an interactive visualisation tool that combines different visualisation methods including tile maps, radial plots and word clouds. By using the visualisation tool with the RockYou database of over 32 million passwords, they discussed different patterns in passwords including dates, e.g., around 5\% of passwords have pure dates, many date-related patterns such as the first days of the month, and holidays were observed. In another follow-up work, Veras et al.~\cite{Veras-R2021} conducted qualitative analyses of leaked password databases using semantic grammar to emanate graphical models for visualising high-level dependencies between token classes. Their work captures both syntactic and semantic information, allowing for the identification of regular patterns in passwords that resemble natural language.

Moreover, researchers have been looking at more subtle password patterns that are less obvious for visual observations. Yu and Liao~\cite{yu2016user} developed a light-weight and web-based visualisation tool combining bar charts, heat maps, tables, and word clouds using the D3 data visualisation library~\cite{michael11} to analyse leaked password databases, which led to the identification of various password patterns (e.g., short and long repeat patterns are common in user passwords, shorter repeating sub-strings are used to form longer repeating sub-strings, and reverse order repetitions are more than forward-order repetitions). In another follow-up work, Yu and Liao~\cite{Yu-Liao2019IJIS} developed hierarchical segmentation and optimisation algorithms to visualise and analyse the prefixes and postfixes of human-created passwords. 

Apart from date-based patterns in human-created passwords and PINs, keyboard-related patterns have also been investigated by some researchers. Schweitzer et al.~\cite{schweitzer09} discovered that drawing lines connecting the key sequences on a graphical keyboard is not good enough to recognise patterns. Alternatively, they developed a new set of rules using (weighted) arcs and/or loops to help visually recognise keyboard patterns. An analysis based on a large number of human-created passwords revealed that the most common keyboard patterns contained 2-4 continuous keys. Based on this result, Chou et al.~\cite{chou2012password} used adjacent and parallel keyboard patterns to generate password databases, and subsequently applied them to crack real-world passwords. 
% MAY CONSIDER TO INCLUDE TWO WEBSITES HERE
% (1) https://wpengine.com/resources/passwords-unmasked-infographic/
% (2) https://informationisbeautiful.net/visualizations/top-500-passwords-visualized/

To the best of our knowledge, there is limited work that is similar to our work presented in this paper. Shin and Woo~\cite{Youjin-S2022} attempted to understand password patterns and structure through a data-driven analysis of passwords from four different leaked password databases. They adopted the tensor decomposition method to study password features and identify two dominant features that make a password stronger through similarity distance analysis. Zheng et al.~\cite{Zheng2018AnAM} proposed a modification-based approach to explore the spatial structure of passwords in the form of entity-relationship graphs. Similar to our work, they also used Levenshtein distance for comparing passwords. However, their approach differed in terms of utilising the Levenshtein distance to define the edges of vertices in a graph model, while we utilise Levenshtein distances between password pairs to generate distance matrices with subsequent dimensionality reduction for mapping complicated spatial password relationships to a 2-D space for visualisation purposes. Guo et al.~\cite{GUO-X2015} also used Levenshtein distances between password pairs to construct a graph showing relationships between passwords, but they used a simple threshold-based approach to define binary connections between passwords, while our work uses a dimensionality reduction method to keep distance between passwords in a 2-D space.

\section{Methodology}
\label{sec:methodology}

%The data used in this study were obtained from the SecLists GitHub repository\footnote{\url{https://github.com/danielmiessler/SecLists}}, specifically from several leaked password data sets that range in size from 12,000\footnote{\url{https://github.com/danielmiessler/SecLists/blob/master/Passwords/Leaked-datasets/bible.txt}} to over 14 million\footnote{\url{https://github.com/danielmiessler/SecLists/blob/master/Passwords/Leaked-datasets/rockyou.txt.tar.gz}}, giving us a variety of leaked dataset sizes we could use. The data sets used contained unique passwords.

The main objective of this work is to develop a tool that facilitates the exploration and analysis of large-scale password databases for researchers and practitioners by leveraging effective data visualisation techniques. To achieve this, we aim to 
\begin{enumerate}
\item construct high-dimensional representations for passwords in a given database, where passwords with similar structures are positioned close together,

\item embed the high-dimensional representations of all passwords in a 2D space, and

\item develop an easy-to-use toolkit for password visualisation and analysis.
\end{enumerate}

%was to develop a visual representation of passwords as points in 2D space, where similar passwords are positioned nearby to each other. 
%This meant that a set of coordinates for each password was required. To do this, we employed the Levensthein distance, a string metric to measure the difference between two sequences, to generate a distance matrix between all passwords, and then used dimensionality reduction to reduce the number of dimensions down to just two.

\subsection{Quantify similarity between a pair of passwords}

The edit distance is a method used to quantify the dissimilarity between two textual strings (e.g., two passwords) by calculating the minimum number of operations needed to transform one string into the other. There are different types of edit distance that involve different sets of editing operations. For instance, Levenshtein distance (hereafter LD)~\cite{Levenshtein-V1966} allows three operations: removal, insertion, and substitution of a character in the input strings. Hamming distance takes effect only on passwords that have the same length. In other words, it does not allow insertion or removal. Jaro-Winkler distance is based on the observation that a common mistake when people type is the transposition of two adjacent characters in a string. It favours strings where the first few characters match due to the prefix scale factor in its calculation (e.g., `password1' and `password2' have a similarity of 95.6\% whereas `1password' and `2password' have a similarity of 92.6\%), but passwords come in many formats and do not necessarily have matching prefixes. Moreover, Jaccard similarity and cosine similarity do not account for the order of characters, which can be critical in comparing passwords~\cite{Bonneau2012}. Cosine similarity also requires a transformation of the strings into a suitable numerical vector representation, which can complicate the process.

Comparing all the different types of edit distances, we selected LD to quantify the similarity between pairs of passwords. The LD between two passwords is formally defined as the minimum number of single-character edits (insertions, deletions, and substitutions) required to change one password into the other. Mathematically, LD between a pair of passwords $a$ and $b$ can be stated by $\operatorname{lev}_{a, b}(|a|,|b|)$:
\begin{equation}
\operatorname{lev}_{a, b}(i, j)=\min \left\{\begin{array}{l}
\operatorname{lev}_{a, b}(i-1, j)+1 \\
\operatorname{lev}_{a, b}(i, j-1)+1 \\
\operatorname{lev}_{a, b}(i-1, j-1)+f\left(a_i, b_j\right)
\end{array}\right.,
\end{equation}
where $|a|$ and $|b|$ represent the lengths of passwords $a$ and $b$, respectively, $f(a_i, b_j)$ is an indicator function that equals to 0 when $a_i = b_j$ and to 1 otherwise. The calculation of LD involves a dynamic optimisation algorithm, whose complexity is $\operatorname{O}(|a|\times|b|)$. 

\begin{table}[!htb]
\centering
\caption{Examples of LDs between three example pairs of passwords}
\label{tab:distance_table}
\begin{tabular}{ccc}
\toprule
\textbf{Password 1} & \textbf{Password 2} & \textbf{LD} \\ 
\midrule
romans56 & blahblah & 8\\ 
bahamut24ritter & Bonito12 & 13\\ 
rahasia23 & abhilash298471 & 11\\
\bottomrule
\end{tabular}
\end{table}

\subsection{Calculating a distance matrix from all passwords}

To facilitate the construction of high-dimensional representations for passwords in a database with respect to other passwords, LDs between all pairs of passwords can be used to create a distance matrix, where each cell represents the similarity between two passwords. In this case, the $i$-th row and the $j$-th column represents the LD between the $i$-th and $j$-th passwords. Table~\ref{tab:distance-matrix} shows an example of a distance matrix of 10 randomly selected passwords from a leaked password database\footnote{\url{https://github.com/danielmiessler/SecLists/blob/master/Passwords/xato-net-10-million-passwords-10000.txt}}.

\begin{table}[!htb]
\centering
\small
\caption{Examples of a distance matrix of a database with 10 passwords}
\label{tab:distance-matrix}
\begin{tabular}{*{11}{c}}
\toprule
& \rotatebox{90}{anfield} & \rotatebox{90}{cutlass} & \rotatebox{90}{denire} & \rotatebox{90}{GEORGE} & \rotatebox{90}{21081987} & \rotatebox{90}{WP2003WP} & \rotatebox{90}{vjqgfhjkm} & \rotatebox{90}{hallo123} & \rotatebox{90}{nathalie} & \rotatebox{90}{november}\\
\midrule
anfield & 0 & 7 & 6 & 7 & 8 & 8 & 8 & 7 & 7 & 7\\
cutlass & 7 & 0 & 7 & 7 & 8 & 8 & 9 & 7 & 6 & 8\\
denire & 6 & 7 & 0 & 6 & 8 & 8 & 9 & 8 & 8 & 7\\
GEORGE & 7 & 7 & 6 & 0 & 8 & 8 & 9 & 8 & 8 & 8\\
21081987 & 8 & 8 & 8 & 8 & 0 & 8 & 9 & 8 & 8 & 8\\
WP2003WP & 8 & 8 & 8 & 8 & 8 & 0 & 9 & 8 & 8 & 8\\
vjqgfhjkm & 8 & 9 & 9 & 9 & 9 & 9 & 0 & 9 & 9 & 9\\
hallo123 & 7 & 7 & 8 & 8 & 8 & 8 & 9 & 0 & 7 & 8\\
nathalie & 7 & 6 & 8 & 8 & 8 & 8 & 9 & 7 & 0 & 7\\
November & 7 & 8 & 7 & 8 & 8 & 8 & 9 & 8 & 7 & 0\\
\bottomrule
\end{tabular}
\end{table}

However, for large leaked password databases, there are too many passwords, so creating a complete distance matrix can incur high time and space complexity. Imagining a best-case scenario for memory usage where we assume that each password is a single character long, amounting to one byte. Even then, a data set with 700,000 passwords would require $700,000 \times 700,000 \times 8 = 3.92$ trillion bits, or 490 gigabytes of memory. Given the high complexity, we resorted to an anchor-based method to make the data visualisation tool more lightweight. The method decided involves selecting a sufficiently small number of representative anchor passwords in the entire database and constructing a distance matrix of all passwords from the anchor passwords only (rather than all). In this way, for a database of size $M$, we can extract a set of $N\ll M$ anchor passwords and generate an $M \times N$ distance matrix, where each row is an $N$-d vector indicating how close (LD-wise) each password is to each of the $N$ anchor passwords. This reduced matrix can be easily accommodated in memory and also enables faster computation in subsequent steps. Crucially, this approach still maintains the variance of the data, providing us with a reliable sample for further analysis.

\subsection{Dimensionality reduction}

In the second step of this process, we used t-SNE~\cite{tsne} to reduce the number of dimensions of each password in the $M\times N$ distance matrix in the previous step from $M>2$ to just two. As an input to the t-SNE method, the high dimensional distance matrix is passed, and the output is a matrix where each row represents the 2-D coordinates for each password. t-SNE is a machine learning algorithm adept at visualising high-dimensional data and preserving the local structure, enabling it to represent clusters and relationships in the data effectively and making it particularly suitable for our goal of producing a two-dimensional representation. One of its key advantages is its ability to maintain the local structure of the data, meaning it preserves the relationships and clusters that exist in the original high-dimensional data. %Given we want to position similar passwords close to each other, a dimensionality reduction method that preserves local structure is ideal and is why t-SNE was chosen.

\subsection{Implementation}

\subsubsection{Python-based command-line tool for visualising large password datasets}
\label{sec:python}

We developed a Python command-line tool to process and visualise large password databases as discussed previously in this section\footnote{\url{https://github.com/samcparker/passviz-cli}}. We chose Python due to its extensive range of data and machine learning libraries. The polyleven library package\footnote{\url{https://ceptord.net/}} was used for calculating LDs. For the t-SNE algorithm, we used the implementation in the openTSNE pacakge\footnote{\url{https://github.com/pavlin-policar/openTSNE}}. The visualisation implementation was done with Matplotlib\footnote{\url{https://matplotlib.org/}} in Python, offering researchers and practitioners various options to interact with the password database to facilitate different follow-up analyses.

\subsubsection{Interactive application}
\label{sec:interactive_application}

In addition, we also developed an interactive web-based application that allows users to explore visualisation of smaller password databases (which can be a smaller subset of a larger password database) in an interactive manner\footnote{\url{https://github.com/samcparker/passviz-gui}}. Here, we offer a glimpse into the capabilities that this interactive application can offer using a smaller password database as an example.

\begin{figure}[!htb]
\centering
\includegraphics[width=0.9\linewidth]{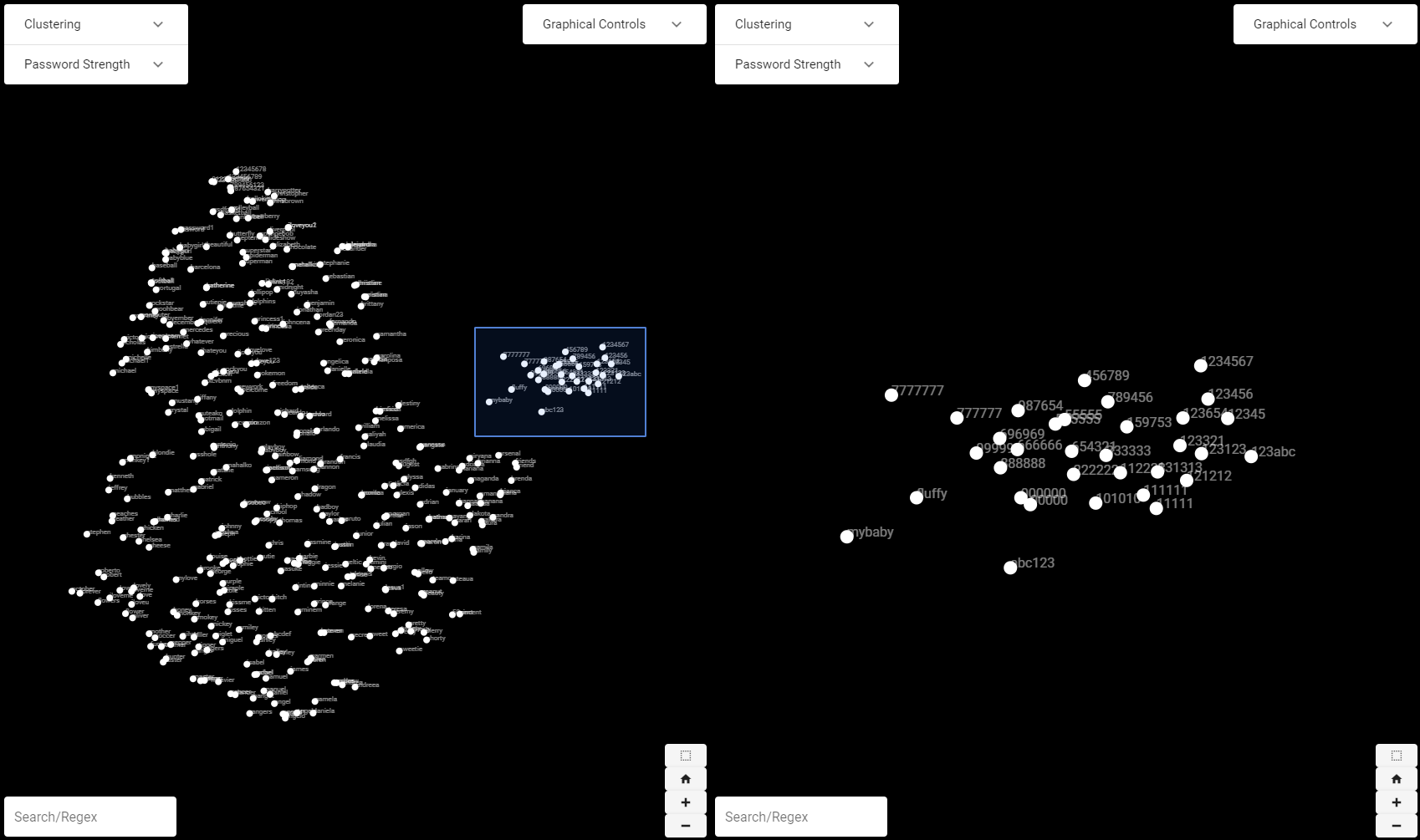}
\caption{Extracting a group of passwords and opening them in a new window in PassViz}
\label{fig:PassViz_extraction}
\end{figure}

\textbf{Extraction}: Figure~\ref{fig:PassViz_extraction} shows the extraction functionality of the application, where a set of passwords can be selected and converted into a new graph for closer inspection. It is possible to take this extracted group of passwords and generate a visualisation of them.

\begin{figure}[!htb]
\centering
\includegraphics[width=0.8\linewidth]{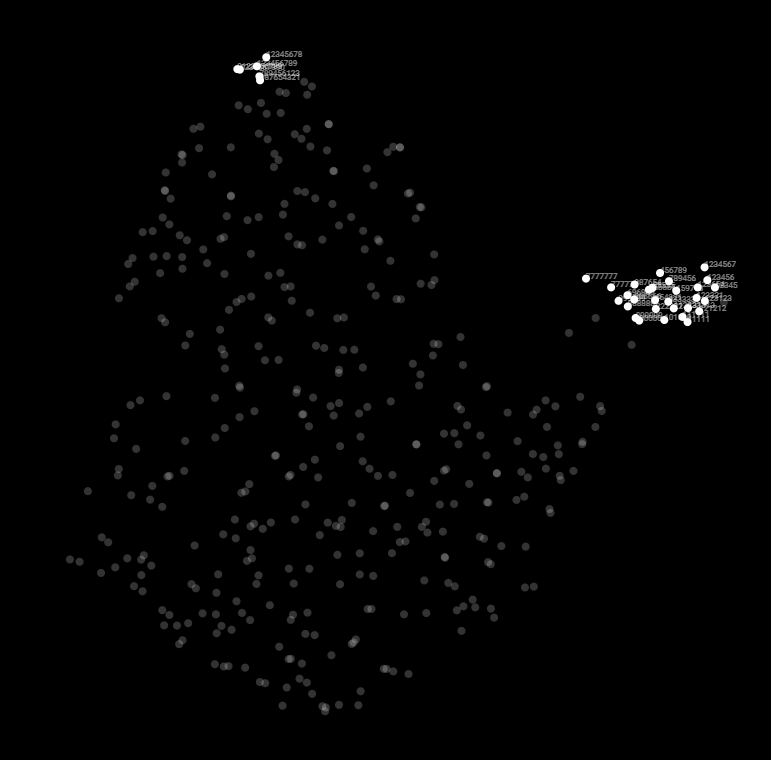}
\caption{Using regular expressions to highlight individual passwords}
\label{fig:PassViz_regex}
\end{figure}

\textbf{Searching}: Using regular expressions, a user is able to hide passwords that do not match the regular expression provided. In Figure~\ref{fig:PassViz_regex}, the regular expression \verb|^[0-9]*$| is used to show only passwords that contain only numbers.

\begin{figure}[!htb]
\centering
\includegraphics[width=0.8\linewidth]{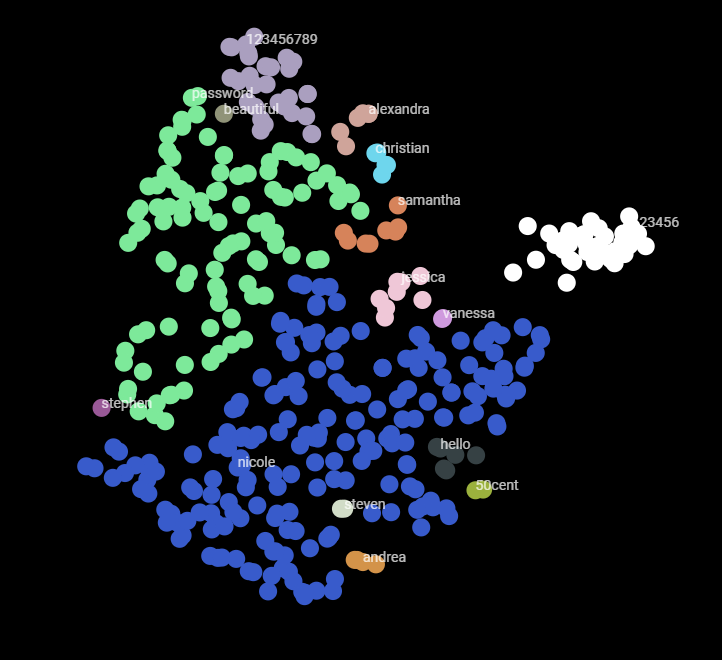}
\caption{Using OPTICS to cluster the passwords}
\label{fig:PassViz_clustering}
\end{figure}

\textbf{Clustering algorithms}: This application has support for performing $k$-means, OPTICS and DBSCAN clustering methods to get a better understanding of where clusters are formed and to allow for easier visualisation of patterns that may not have emerged before. Figure~\ref{fig:PassViz_clustering} shows the graph after having the OPTICS clustering method performed on it. By performing this clustering algorithm, the application highlights the centre-most passwords within the clusters. This reveals `andrea' to be the centre of the cluster containing passwords `andres', `andrew', `andrea' and `andreea'.

% Maybe include some stuff on UMAP here. I feel both are good, so maybe for the final paper we could show the difference in results, or change our methodology to say we didn't chose one of them for certain reasons we found in our end results (e.g. t-SNE was too sparse).

Due to the time and space complexity of processing large password databases from a web browser, this interactive application will be very slow or even impossible to load and process very large password databases. Therefore, we recommend using this interactive application as a complementary tool alongside the command-line tool that is better positioned to process and visualise large-scale password databases. This combination enables us to delve into interesting subsets of a large password database to study more hidden patterns, therefore enhancing insights and findings learned from the results from the large database. The interactive application can also be used to test some hypotheses with a small subset of a large password database, and then a more time-consuming process is run using the command-line tool with the full password database. Examples of various password analyses are presented in Section~\ref{sec:results} using a leaked database `000webhost'. Nevertheless, one major direction of our future work is to investigate how the time and space complexity of this interactive application can be improved to handle larger password databases, e.g., leveraging parallel processing using multiple cloud servers and GPUs on a single machine.

\section{Experimental Results}
\label{sec:results}

\begin{figure}[!htb]
\centering
\includegraphics[width=.7\linewidth]{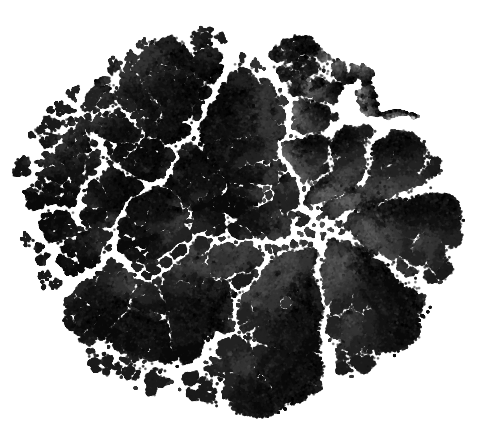}
\caption{Illustration of clusters of passwords}
\label{fig:passwords_cluters}
\end{figure}

In this section, we present our work of applying PassViz to analyse passwords in the leaked database `000webhost' to showcase its capabilities and evaluate its effectiveness. 000webhost comprises 15,251,074 clear text passwords, including 720,302 unique passwords. This leaked password database was made public in November 2015, following a security breach from a large web hosting service 000webhost.com. According to a study~\cite{Veras-R2021}, the origin of the user accounts in 000webhost is reported to be diverse. All accounts in this database are distributed across a wide range of countries, where the largest one (United States) only accounts for 8\% of the total population. In addition, the distribution indicates that English passwords are not dominant in the database~\cite{Veras-R2021}. By using a subset of randomly selected passwords with the size of 2,000, we were able to construct a distance matrix with the size of $720,302 \times 2,000$. After applying t-SNE dimensionality reduction, we were able to plot all passwords as a 2-D graph and show them in different clusters. As illustrated in Figure~\ref{fig:passwords_cluters}, PassViz could group all passwords into discernible clusters. To further learn and understand more patterns in this leaked database, more analyses were performed and the findings are presented below.

\subsection{Analysis based on password length}
\label{sec:analysis_length}

As shown in Figure~\ref{fig:passwords_cluters}, different clusters can be visually observed. However, it is not clear what the most defining factor of the clusters is. By looking into the subsets of the database through the utilisation of the interactive application introduced in Section~\ref{sec:interactive_application}, these clusters are primarily differentiated by the length of the passwords. We performed a number of analyses to further explore the impact of password length on the visualisation of the password database. 

\subsubsection{Clusters based on different password lengths}

\begin{figure}[!htb]
\centering
\includegraphics[width=.7\linewidth]{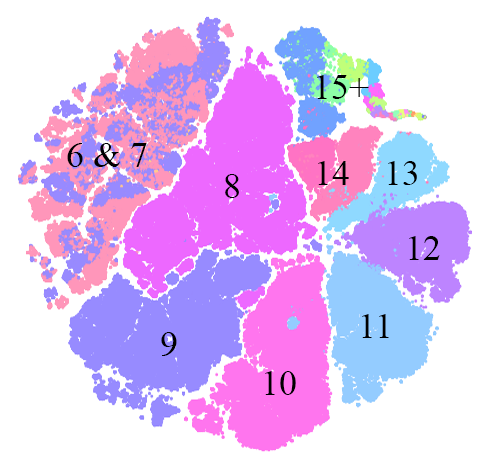}
\caption{Illustration of colour-coded clusters based on different password lengths}
\label{fig:passwords_cluters_length}
\end{figure}

We conducted a subsequent analysis to encode different password lengths with different colours for visualisation. As shown in Figure~\ref{fig:passwords_cluters_length}, the visualised database illustrates each password in a colour corresponding to its length. Additionally, a number displayed over each cluster indicates the majority length of the passwords contained within. The size of a cluster corresponds to the number of passwords that have the same length. It is apparent from this method of visualisation that the length of the password is a significant factor in the formation of the clusters, showing that the length of passwords plays a significant role in the structure of passwords within the database.

However, there are exceptions to this observation. One instance is the formation of a mixed cluster, predominantly consisting of passwords of lengths 6 and 7. Despite the minor difference in length, these passwords have enough in common to be grouped into the same cluster. Another exception to this observation is for passwords that have 15 or more characters. Rather than forming individual clusters, these longer passwords combine into a single cluster. The group gradually gets smaller as the length of the passwords increases, reflecting fewer instances of longer passwords in the database. Moreover, it was observed that no cluster contained passwords with fewer than 6 characters, suggesting that 000webhost might have enforced a password-composition policy to have a minimum password length of 6 characters. The identification of larger clusters of passwords with lengths of 8, 9, and 10 is somewhat consistent with the findings reported in~\cite{Komanduri-S2011}, which revealed that password-composition policies mandating a minimum of 8 characters typically result in mean password lengths ranging from 9 to 10.

\subsubsection{Visualising passwords of the same length}

\begin{figure}[!htb]
\centering
\includegraphics[width=.8\linewidth]{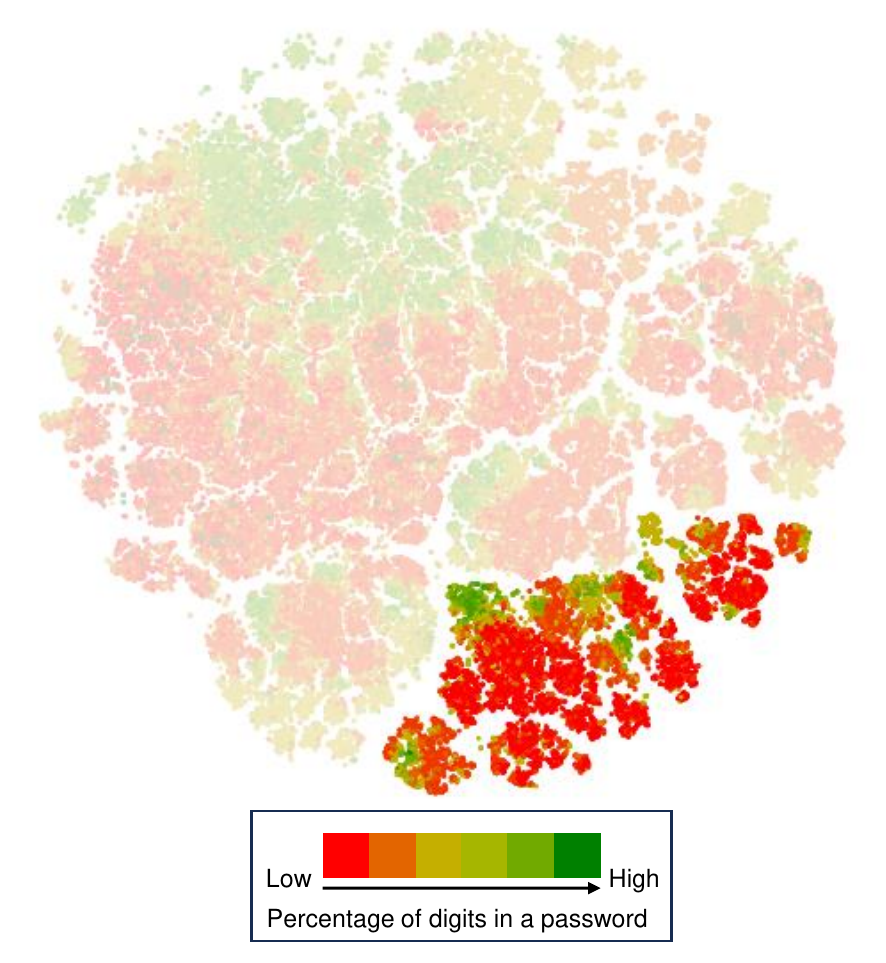}
\caption{Visualisation of passwords of the length of 8 characters}
\label{fig:passwords_cluters_length_8}
\end{figure}

From the previous analysis, it is worth noticing that the defining factor separating leaked password databases into clusters is the length of the password. To further explore, we take passwords of the length of 8 as an example to illustrate how the 000webhost database graph transforms. Around 140,000 passwords are plotted as shown in Figure~\ref{fig:passwords_cluters_length_8}, which gives a better understanding of how graphs are formed, without the aforementioned bias of length down to Levenshtein distance.

\begin{figure}[!htb]
\centering
\includegraphics[width=.8\linewidth]{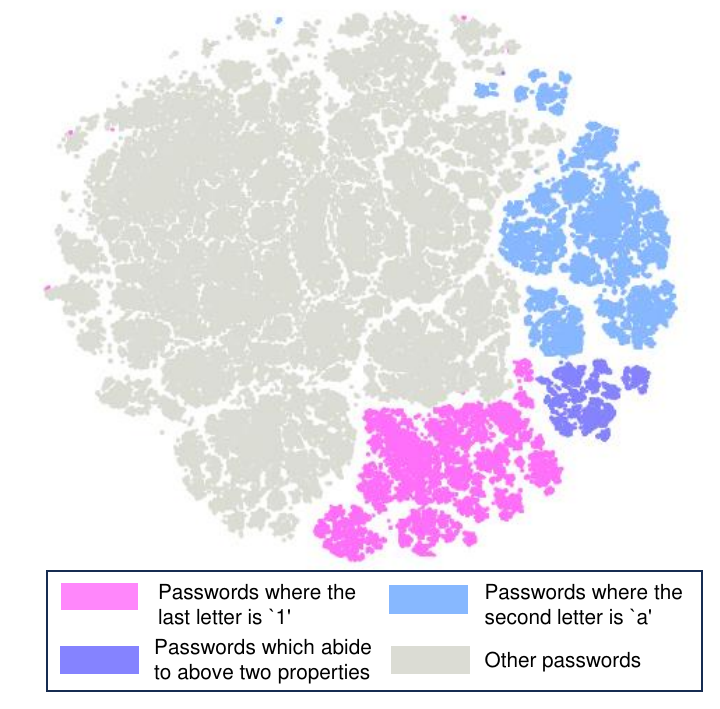}
\caption{Visualisation of passwords of length 8 that have specific compositions}
\label{fig:passwords_specific_composition}
\end{figure}

By visualising the passwords in this way and after further analysis, a common pattern emerged in that many of the passwords in certain clusters had the same character at the same position in each password. In Figure~\ref{fig:passwords_specific_composition}, each password has been given a colour based on a property: blue represents passwords where the second letter is `a', pink represents passwords where the last letter is `1', and purple represents passwords that abide by both of these properties. It seems that the most defining factor of passwords in our methodology is the position of characters within the passwords. It is striking to see how many users include `a' as the second letter of their password or `1' as the last letter and is a pattern that would be harder to visualise using more common statistical methods.

%As previously shown on an entire data set as a whole, the following image visualises how the generated graph looks where red-coloured passwords represent passwords with a low percentage of numbers and green-coloured passwords represent passwords with a high percentage of numbers. Again, we can see some form of gradient from one side of the password to the other, with some clusters appearing very dark and some clusters appearing a bit lighter.
%\includegraphics[width=\linewidth]{figures/colour percentage length 8.png}

In addition, looking closely at passwords ending with `1', by isolating the cluster, it appears that many of these passwords only contain a small amount of digits, shown by the majority of passwords appearing red as shown in Figure~\ref{fig:passwords_cluters_length_8}. In comparison, other clusters have more of an orange-to-green hue, showing they contain more numbers.

\subsection{Analysis based on the composition of digits in a password} 
\label{sec:analysis_digits}

Many existing works have looked into the composition of a password~\cite{Komanduri-S2011}. How digits play a role in creating a password is often of interest to researchers and practitioners. Here, we present a number of analyses utilising PassViz to help derive insights from the large-scale password database by looking at the composition of digits within the password. 

\subsubsection{Visualisation based on the percentage of digits in a password}

\begin{figure}[!htb]
\centering
\includegraphics[width=.8\linewidth]{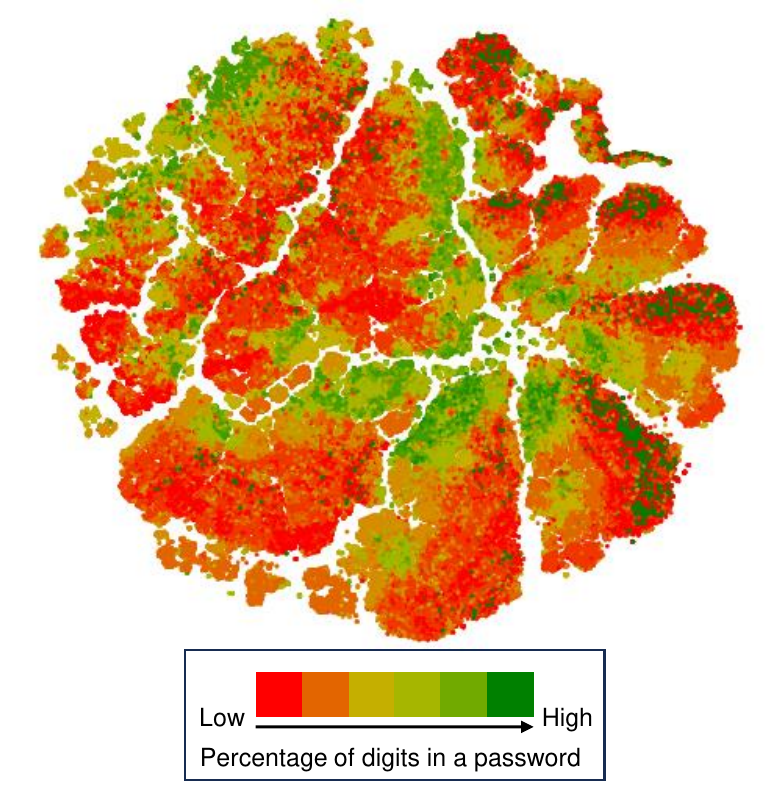}
\caption{Visualisation of passwords that have different percentages of numbers}
\label{fig:passwords_percentage_number}
\end{figure}

Upon assessing the numerical composition of passwords, they were colour-coded based on the percentage of digits in a password, where passwords with dark green colour have the highest percentage of digits and passwords with dark red colour have the lowest percentage of digits. % with passwords resembling a colour between red and green based on the percentage of numbers in the password, with red passwords representing passwords with few numbers and green passwords representing passwords with lots of numbers. 
As shown in Figure~\ref{fig:passwords_percentage_number}, the visualisation revealed that the utilisation of the numerical composition of a password can separate passwords within their clusters, with one side of the cluster containing passwords having a high percentage of digits, and the other side containing passwords with fewer digits. There appears to be a gradient across all clusters, visualising the change in the number of digits contained within passwords.

To facilitate the analysis, we present Table~\ref{tab:table_percentage_number} which displays the distribution of numerical composition in the 000webhost database. The table reveals that 21\% of passwords in the database contain 20\% digits, while 17\% and 16\% of passwords have 10\% and 30\% digits, respectively. Including 5\% of passwords that have no digits, 59\& of passwords in the 000webhost database have less than 30\% numerical content, which makes the overall graph lean towards the colour of red.

\begin{table}[!htb]
\centering
\small
\caption{Distribution of the percentage of digits in each password in the 000webhost database}
\label{tab:table_percentage_number}
\begin{tabular}{cccc}
\toprule
\textbf{\% digits} & \textbf{\% passwords} & \textbf{\% digits cont.} & \textbf{\% passwords cont.}\\
\midrule
0 & 5 & 60 & 7\\
10 & 17 & 70 & 4\\
20 & 21 & 80 & 4\\
30 & 16 & 90 & 0.7\\
40 & 10 & 100 & 0.06\\
50 & 13 & - & -\\
\bottomrule
\end{tabular}
\end{table}

%In addition, we produce a line chart depicted in Figure~\ref{fig:plot_percentage_numbers} to show the number of passwords with a percentage of numbers in the password. It shows that the majority of passwords have 20\% numbers, which clarifies why the image above appears so red.

\iffalse
\begin{figure}[!htb]
\begin{tikzpicture}
\begin{axis} [ybar,
    bar width = 4pt,
    xtick = data,
    ylabel={Number of passwords},
    xlabel={Percentage of numbers in password}
]
\addplot [smooth] plot coordinates {
(0, 37388)
(10, 125823)
(20, 151713)
(30, 114038)
(40, 70615)
(50, 94858)
(60, 52907)
(70, 33883)
(80, 33331)
(90, 5257)
(100, 478)
};
\end{axis}
\end{tikzpicture}
\caption{Distribution of percentage of numbers for all passwords in the 000webhost password database} 
\label{fig:plot_percentage_numbers}
\end{figure}
\fi

\subsubsection{Visualisation based on the dominating position of digits}

\begin{figure}[!htb]
\centering
\includegraphics[width=.8\linewidth]{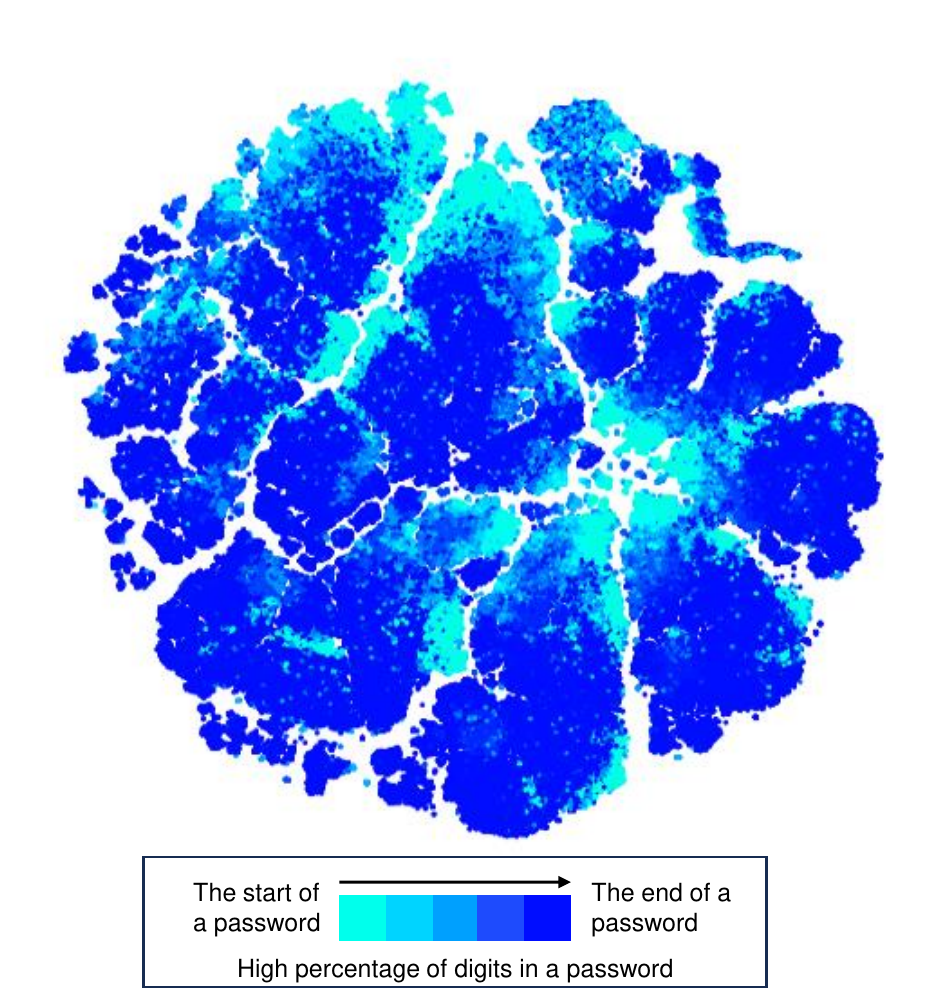}
\caption{Visualisation of positional distribution of digits within a password}
\label{fig:passwords_digits_distribution}
\end{figure}

An interesting property to be examined is the positional distribution of digits within a password. This metric measures the location of digits within a password, where values near 0 signify that numbers are primarily concentrated towards the beginning of the password, and values closer to 1 indicate numerical characters mostly towards the end. A distribution ratio around 0.5 suggests either an even dispersion of digits or a lack of digits altogether. Figure~\ref{fig:passwords_digits_distribution} depicts this metric, with the light blue shaded passwords indicating a higher quantity of digits towards the start of the passwords and the dark blue passwords signifying a greater presence of digits towards the end. %This factor does not appear to have as much of an effect on the clustering as the password length of proportion of numerical characters, but it does form distinguishable patterns within the clusters.
A noteworthy observation is the comparative scarcity of passwords with a high predominance of digits towards the start as opposed to passwords with a majority of digits towards the end. This implies a bias towards appending numbers at the end of the passwords.

\subsection{Analysis based on specific requirements}
\label{sec:analysis_spec}

To assist the exploration of a large-scale password database to provide more insights, PassViz has the capability and flexibility to produce visualisation based on more specific requirements. Here we present some examples of utilising PassViz to learn password patterns and structures. 

\subsubsection{Visualisation based on a given string}

\begin{figure}[!htb]
\centering
\includegraphics[width=.9\linewidth]{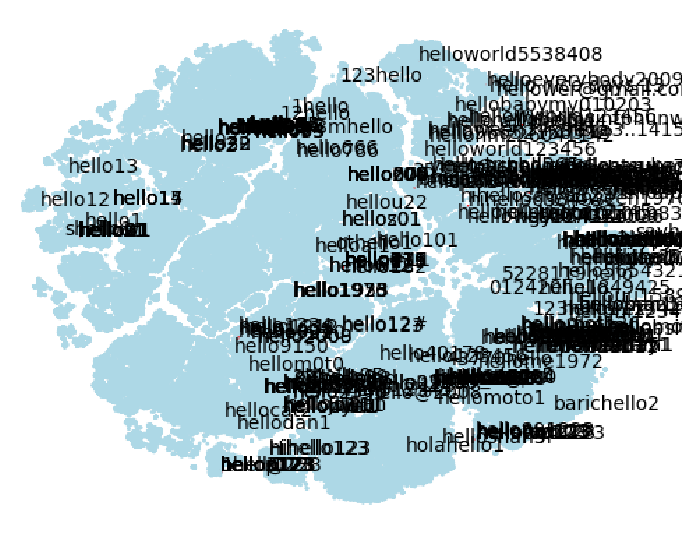}
\caption{Visualisation of passwords that contain the word `hello'}
\label{fig:passwords_hello}
\end{figure}

All instances containing the word `hello' in the 000webhost database were highlighted as shown in Figure~\ref{fig:passwords_hello}. These instances are relatively scarce, however, there are occasionally groupings. This suggests that while the strings may appear similar based on their contents, it is not a decisive factor in global cluster formation, only in local formation within clusters.

\subsubsection{Visualisation based on passwords containing years}

\begin{figure}[!htb]
\centering
\includegraphics[width=.8\linewidth]{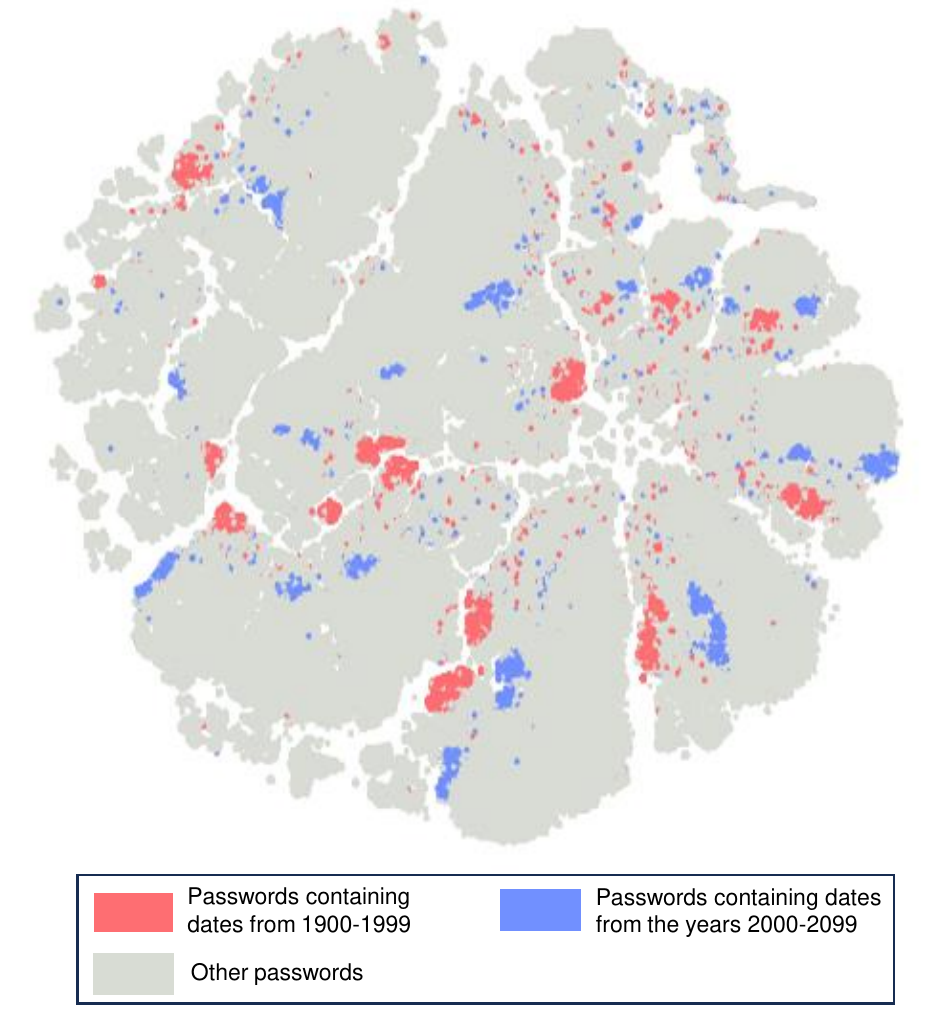}
\caption{Visualisation of passwords that contain years}
\label{fig:passwords_years}
\end{figure}

In addition, existing works discovered that numbers that represent dates/years have been frequently used in the process of password/PIN creation~\cite{Wang2017AsiaCCS, Veras2012VizSec}. We are interested in visualisation the distribution of such information in the 000webhost database using PassViz to see if there are any interesting patterns that can be discovered. As depicted in Figure~\ref{fig:passwords_years}, passwords containing dates from the years 2000-2099 are highlighted in blue, such as `amado2009', while those containing dates from 1900-1999, like `small1970sman', are marked in red. These dates were chosen specifically as they have the most relevance to current users.

In this visualisation, it can be seen that a small portion of passwords containing these year-related numbers are scattered across the graph. The larger clusters in blue and red formed primarily consist of passwords where the year forms the end part of the password, like `amado2009', suggesting that a substantial portion of users with a date in their password append a specific year to a base word, rather than at the start or in the middle of the password, contributing to the formation of these clusters. On the other hand, the individual points are scattered throughout the clusters representing the less common instances where the year appears in the middle of a password such as `small1970sman', rather than at the end of the password. 

\subsection{Comparative analysis}
\label{sec:analysis_compare}

\subsubsection{Comparing 000webhost with phpbb based on the percentage of digits}

\begin{figure}[!htb]
\centering
\includegraphics[width=.8\linewidth]{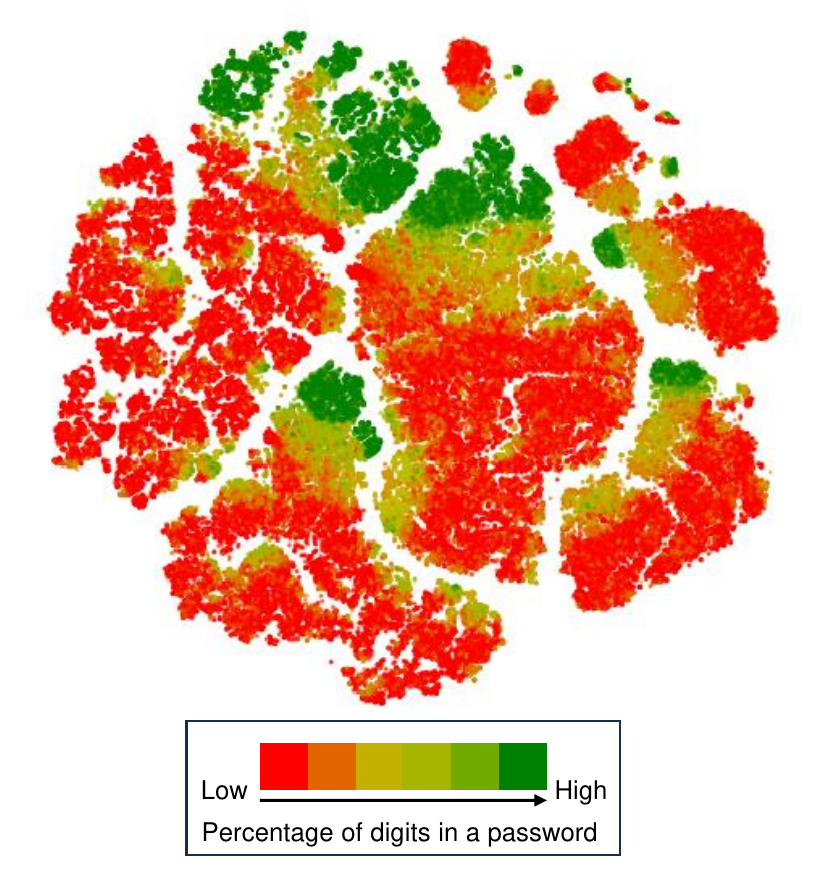}
\caption{Visualisation of passwords that have different percentages of digits in the phpbb database}
\label{fig:phpbb_colour_percentage}
\end{figure}

This generation methodology in this research can be extended and applied to different databases. To illustrate this, we present the graph for the leaked database `phpbb'\footnote{\url{https://github.com/danielmiessler/SecLists/blob/master/Passwords/Leaked-Databases/phpbb.txt}} of over 184,000 unique passwords, shown in Figure~\ref{fig:phpbb_colour_percentage}. This shows the percentage of digits in passwords, with red indicating passwords containing no digits, green indicating passwords containing only digits, and colours in-between showing a colour in-between showing the proportion of digits in the password.

Comparing this with Figure~\ref{fig:passwords_percentage_number}, an intriguing pattern becomes evident. The proportion of passwords in the phpbb database containing mostly digits is notably larger compared to the 000webhost database shown in Figure~\ref{fig:passwords_percentage_number}. Additionally, the significant amount of green highlights the prevalence of passwords composed exclusively of digits.

This gives us an insight into the general security of passwords in each database. The graph generated using the 000webhost database does not show the intense red or intense green that the phpbb database shows, indicating that the passwords used within it are more secure. This could be down to security restrictions imposed on users requiring them to use digits in their passwords. On the other hand, the phpbb database generates a predominantly red and green graph, with only a small amount of colour in between. This indicates that the security restrictions imposed on users were not the same as intense as on 000webhost.

\subsubsection{Comparing 000webhost with phpbb based on sequences}

In this section, we focus on the prevalence of numeric sequences and keyboard patterns in the passwords in each database. For this analysis, a visual representation was generated where the passwords containing numeric sequences (such as `123' and `1234') are marked in red and those satisfying keyboard passwords (consecutive keyboard entries, such as `qwerty' and `zxcvb') are marked in blue.

\begin{figure}[!htb]
\centering
\includegraphics[width=.8\linewidth]{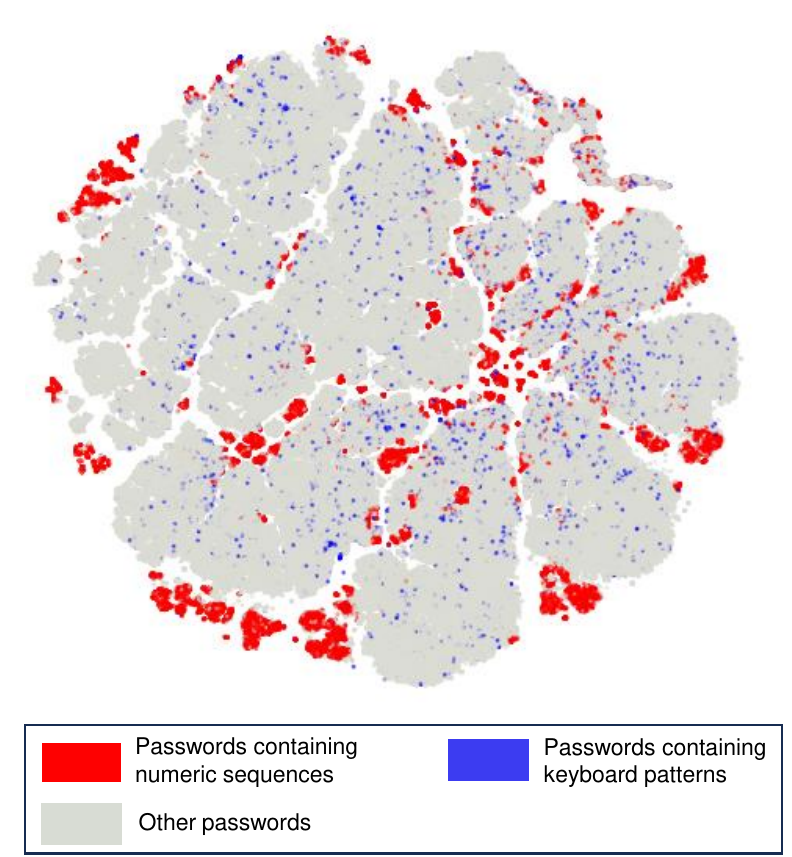}
\caption{Sequences highlighted in the 000webhost database}
\label{fig:000webhost_patterns}
\end{figure}

The 000webhost database illustrated a pronounced prevalence of numeric sequence patterns in password creation, as shown in Figure~\ref{fig:000webhost_patterns}. A substantial proportion of the passwords contained easily identifiable sequences which start with `123' and increase incrementally. This `123' pattern, despite being a weak password strategy, is in much use among passwords in the 000webhost database.

\begin{figure}[!htb]
\centering
\includegraphics[width=.9\linewidth]{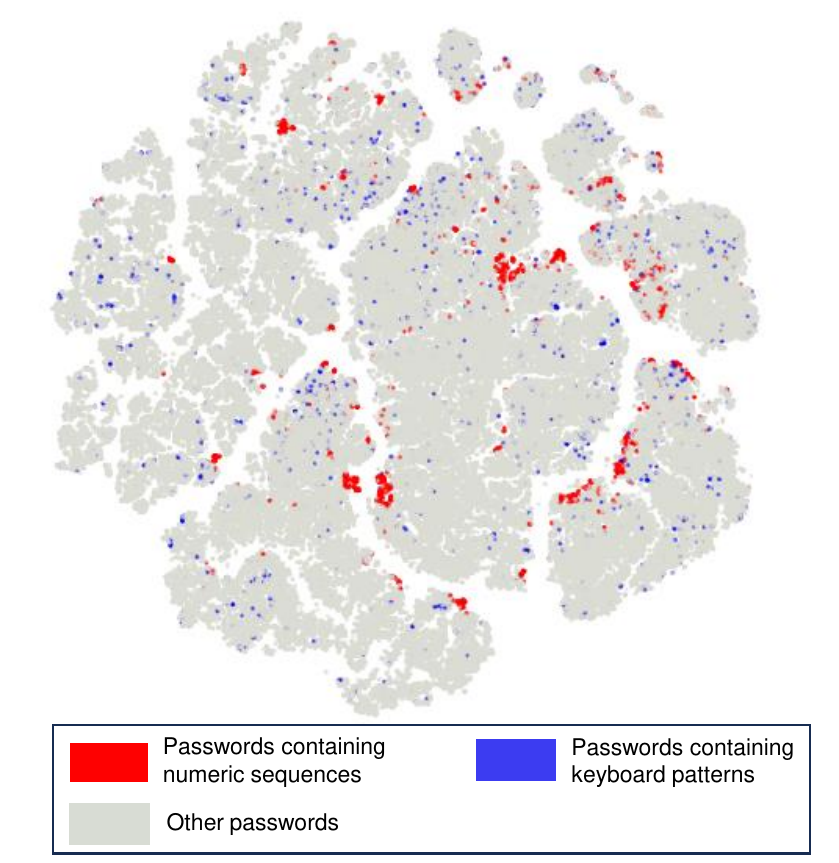}
\caption{Sequences highlighted in the phpbb database}
\label{fig:phpbb_patterns}
\end{figure}

On the contrary, the phpbb database demonstrated significantly fewer instances of numeric sequence patterns, as shown in Figure~\ref{fig:phpbb_patterns}. This disparity suggests that phpbb users may have had a better understanding of secure passwords, or that the platform itself may have enforced stricter password policies. However, when comparing it with Figure~\ref{fig:phpbb_colour_percentage}, we can see that few passwords in the phpbb database contain digits, which is likely the reason why the `123' pattern is not common inside of the phpbb database.

\subsubsection{Intersection between 000webhost and phpbb}

\begin{figure}[!htb]
\centering
\includegraphics[width=.8\linewidth]{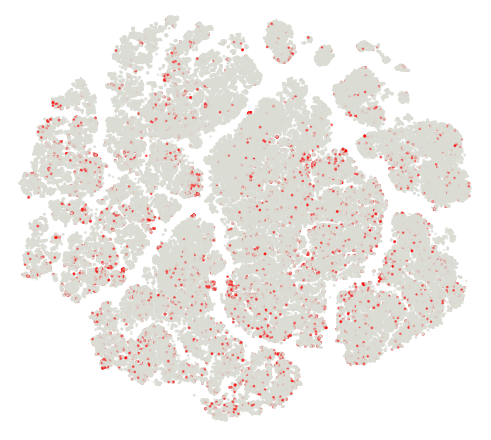}
\caption{Passwords in the phpbb database, with red dots representing passwords shared with the 000webhost databbase}
\label{fig:000webhost_phpbbintersection}
\end{figure}

Calculating the intersection between two password databases may indicate the similarities between them and will show common passwords between the two. Figure~\ref{fig:000webhost_phpbbintersection} visualises passwords in the phpbb database, with passwords marked in red that also appear within the 000webhost database. The number of intersections between the two databases is 6,091 -- ~0.84\% of 000webhost and ~3.3\% of phpbb -- showing a small amount of commonality between the two databases. This is an important area to look at, as it shows where passwords are being re-used between the two databases and will highlight instances where users are re-using seemingly unique passwords across multiple platforms.

In the figure, it can be seen that intersecting passwords are distributed in a non-uniform manner. There are some areas in the graph where marked passwords have a higher concentration, highlighting that there are certain types of passwords that are more likely to be re-used. After further examination on these groups of passwords, it can be seen that these concentrated areas are formed of passwords ending with the characters of `123'.

Some notable instances are longer passwords that may appear random. By performing this method of visualisation, we can see that they are re-used across both databases, despite appearing seemingly unique. These passwords typically appear at longer lengths and include `richmond1969', `serkan737526' and `nikita040683'. One could expect that these come from the same user having created accounts on both platforms.

There are other longer passwords that appear in both databases, however, these do not appear to be as random and unique as the previous one. These come in the form `1q2w3e4r5t6y', `abc123def456' and `qwe123asd456'. Despite being long, these passwords are not necessarily unique and are formed of common keyboard patterns. Thus, using a long set of characters does not necessarily imply a unique password.

\begin{figure}[!htb]
\centering
\includegraphics[width=.9\linewidth]{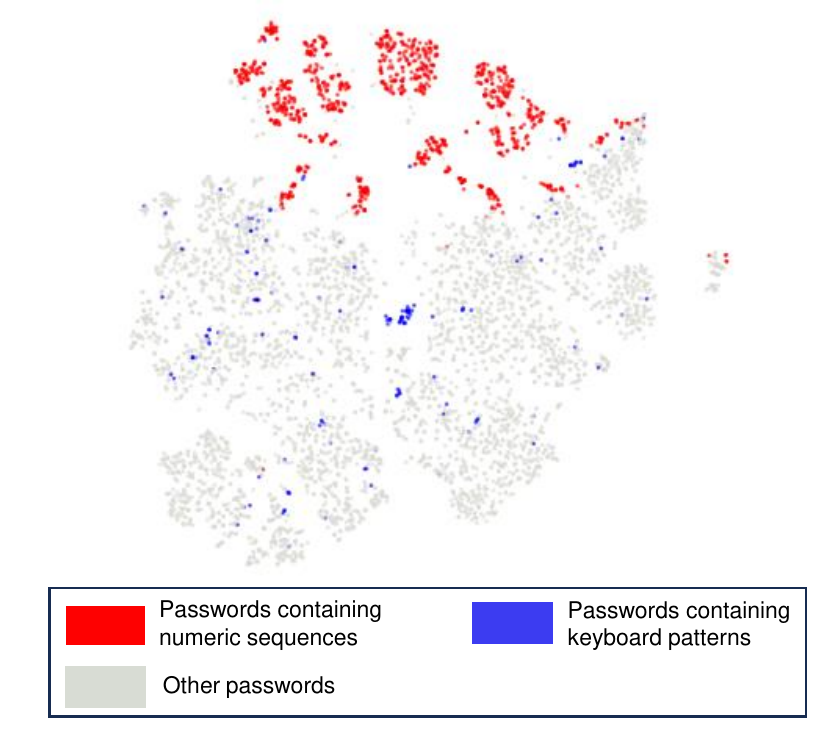}
\caption{Intersection between the 000webhost and phpbb databases}
\label{fig:000webhost_phpbbintersection_graph}
\end{figure}

To get a better understanding of the intersection between these two databases, Figure~\ref{fig:000webhost_phpbbintersection_graph} visualises the graph generated by the intersection of the 000webhost and phpbb databases. As shown in Figure~\ref{fig:000webhost_patterns}, passwords that contain a numeric sequence are highlighted in red, and passwords that contain keyboard sequences are highlighted in blue. This graph reinforces how many passwords re-used between the two databases contain numeric sequences, as this is one of the most defining features of this generated graph. Other notable areas are the reuse of passwords containing other keyboard sequences, like `qwerty', which also appear frequently throughout the graph.

\section{Further Discussions}

\subsection{Typical use cases}

In this subsection, we list some typical use cases of PassViz in real-world applications, based on the examples explained in the previous section. Note that the list is not exhaustive.

\begin{itemize}
\item Use Case 1: As presented in Sections~\ref{sec:analysis_length} and \ref{sec:analysis_digits}, PassViz can be used to reinforce the comprehension of password structures and patterns, thereby extracting valuable insights about human users' password creation processes. In turn, this will aid in the refinement and development of password tools like password strength meters and password policies.  
    
\item Use Case 2: As demonstrated in Section~\ref{sec:analysis_spec}, PassViz provides flexible ways to allow researchers and practitioners to interact with a large password database with ease to explore finer structures related to subsets. 
    
\item Use Case 3: As illustrated in Section~\ref{sec:analysis_compare}, PassViz allows the comparison between two password databases, in order to reveal cross-database/website patterns that cannot be revealed by studying the multiple password databases separately.

\item Use Case 4: All the analyses supported by PassViz can help unveil different aspects of human behaviours in the password creation process, e.g., how they use numbers and keyboard patterns, how they apply character transformation rules to make a password more complicated, and how they reuse or change behaviours across different websites. Such insights related to human behaviours can be useful for a wide range of applications, including development of better tools and also better ways to educate users about password security.
\end{itemize}

\subsection{Limitations}

Despite their utility, the generated visualisations do show certain limitations. We discuss some of such limitations below.

\begin{figure}[!htb]
\centering
\includegraphics[width=.7\linewidth]{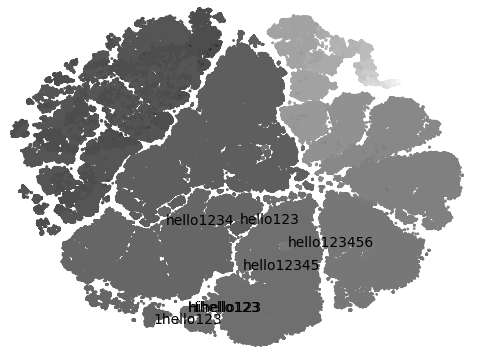}
\caption{An example of one observed limitation of PassViz for `hello123' and `hello12'}
\label{fig:passwords_limitation1}
\end{figure}

One limitation is that some passwords with a small LD could be mis-clustered. For instance, passwords such as `hello123' and `hello12', while notably similar with a LD of 1, are segregated as shown in Figure~\ref{fig:passwords_limitation1}. It is likely due to bias towards strings of identical length within a distance matrix. Unavoidable errors introduced by the dimensionality reduction algorithm may be another source.

\begin{figure}[!htb]
\centering
\includegraphics[width=.7\linewidth]{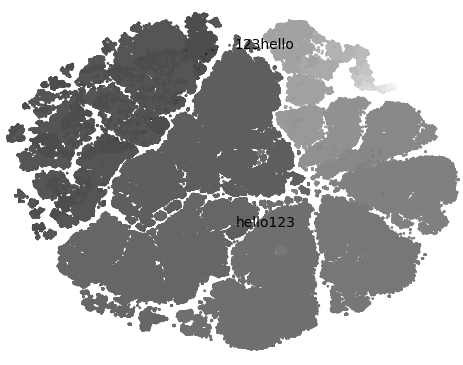}
\caption{An example of a second observed limitation of PassViz for `hello123' and `123hello'}
\label{fig:passwords_limitation2}
\end{figure}

A second limitation is that the LD we used is unable to capture all aspects of semantic similarity of two passwords. For instance, `hello123' and `123hello', despite their perceptible semantic similarity, appear substantially distanced from each other as illustrated in Figure~\ref{fig:passwords_limitation2}. Considering their LD is indeed relatively large (5), the separation can be conceptually explained by the limitation of LD as an edit distance without considering reversing the whole string as a single editing step.
%However, this divergence could also be attributed to the characteristics of the clustering algorithm employed, which aims to preserve the relative distances between all points in the database while reducing dimensions.

A third limitation is that the use of a dimensional reduction algorithm will unavoidably lead to loss of information for some password pairs (so their distances can be more distorted than others). How to address this is non-trivial since we have to visualise passwords in a low-dimensional (2-D or 3-D) space.

\subsection{Future Work}

We have identified a range of future work directions as described below.

\textbf{More comprehensive testing and password analyses:}
The password analyses we conducted and reported are relatively ad hoc, and we only tested PassViz with a number of leaked password databases and some patterns. It will be useful to conduct a more comprehensive analysis with more leaked password databases and a more comprehensive set of patterns. It will also be helpful to design ways to reveal more unknown patterns about passwords. Such further analyses can also involve recruitment of human participants to more confidently confirm the usefulness of PassViz.

\textbf{Refinement our methodology:}
The current study uses a distance matrix based on LD for password position generation. While this approach has its merits, it may not necessarily provide the most comprehensive or meaningful results. Future work could focus on exploring alternative distances, dimensional reduction and clustering methods, such as term frequency $n$-grams and other semantic analysis based vectorisation methods, which could potentially show patterns in passwords that might be overlooked or mis-handled by LD. One interesting approach is to use a large language model (LLM) to define a more semantically aware distance and use the LLM to select more representative anchor passwords of the whole password space. Incorporating a password strength meter in the distance metric may also be useful. We also plan to enhance the reconfigurability of PassViz to support different distance metrics and different dimensional reduction methods.

\textbf{Further development of visualisation tool:}
An additional practical implication of this research is the further development of a more comprehensive application that implements more features required to visualise and analyse leaked passwords, especially allowing interactive visual analytics of large password databases.

\textbf{Integration of chatbots:}
One prospect for the future would be the integration of a chatbot capable of interpreting both graphical input and creating command-line actions. This tool could generate, visualise and analyse graphs autonomously, streamlining the process and reducing the manual workload. With advancements in natural language processing especially LLMs, the development and implementation of such chatbots has become increasingly feasible.

\section{Conclusion}
\label{sec:conclusion}

In conclusion, the work we have produced has provided visual insights into the underlying passwords and structures within large-scale password databases. It has shown a multitude of patterns and correlations that might have remained hidden in traditional statistical analyses. These visualisations lie not only in their ability to summarise complex databases, but also have the potential to inform password security policies and user education efforts.

\bibliographystyle{abbrv-doi}
\bibliography{main}

\end{document}